\documentclass[12pt]{article}

\usepackage{graphicx}

\begin{document}
\baselineskip 24.8pt
\hfill Applied Physics Report: 2004--23

\vspace{2\parskip}
\begin{center}
\textbf{\large Thermal transport in SiC nanostructures}

\vspace{1\parskip}
E. Ziambaras\footnote{Corresponding author: Tel. +46 31 772 8427;
Fax: +46 31 772 8426; E-mail: \texttt{eleni@fy.chalmers.se}}
 and P. Hyldgaard

Department of Applied Physics, Chalmers University of Technology and
G\"{o}teborg University, SE-412 96 G\"{o}teborg, Sweden\\
(June 22, 2004)
\end{center}

\vspace{\parskip}\textbf{Abstract}

SiC is a robust semiconductor material considered ideal for
high-power application due to its material stability and large
bulk thermal conductivity defined by the very fast phonons.
In this paper, however, we show that both material-interface
scattering and total-internal reflection significantly
limit the SiC-nanostructure phonon transport and hence the
heat dissipation in a typical device. For simplicity we 
focus on planar SiC nanostructures and calculate the thermal 
transport both parallel to the layers in a substrate/SiC/oxide
heterostructure and across a SiC/metal gate or contact.  
We find that the phonon-interface scattering produces a
heterostructure thermal conductivity significantly smaller than
what is predicted in a traditional heat-transport calculation.
We also document that the high-temperature heat flow across 
the metal/SiC interface is limited by total-internal reflection 
effects and maximizes with a small difference in the
metal/SiC sound velocities.


\vspace{\parskip}\textbf{1. Introduction}

SiC is an exciting material with possible application
in high-power devices but also posing interesting challenges
for our understanding of the necessary heat dissipation.
There are several SiC advantages, for example, structural 
robustness, presence of a native oxide (SiO$_2$), durability, 
and ability to grow high-mobility SiC layers on top of cheap 
Si substrates. The set of very fast phonons,  the high purity, 
and the long phonon mean free path $l_{\rm mfp}$ also ensures 
a very large bulk thermal conductivity.  On the one hand, this 
raises the promise for high-power device applications since it 
indicates an effective heat dissipation. On the other hand, 
the micron scale $l_{\rm mfp}$ invalidates the foundation of 
the traditional analysis based on the Fourier law of thermal 
transport. A finite-element approach that describes the
local flow in terms of bulk values for the thermal 
conductivities breaks down because the phonon mean free 
path now approaches or exceeds typical feature sizes 
(thickness of conduction channel and oxide layers and 
distance between gates and contacts).  We must instead turn 
to phonon-based calculations to describe SiC devices that 
represent a natural test case for developing an understanding 
of nanostructure thermal transport~\cite{cahill}.

Figure 1 shows a structural schematics (upper panel)
of a SiC device for which we document two phonon transport 
effects (lower panels) that significantly suppresses the heat 
dissipation. We observe that the thermal transport is completely
dominated by the phonon dynamics even when discussing the 
conductance across a SiC/metal interface. For the phonon
transmission problem we calculate the thermal conductance 
of SiC interfaces at room temperature, $\theta=\hbox{300 K}$, 
and report the variation with general ratios of the material 
acoustic impedances and sound velocities.  We find that this 
thermal conductance exhibits only a weak dependence on the 
ratio of acoustic impedances (which determines the long-wavelength 
phonon transmission) but a strong variation with the ratio of 
the material sound velocities (that define the total internal 
reflection of phonons).  In contrast to what is normally assumed, 
we argue that heat dissipation can be maximized by matching
the sound velocities of the SiC and metal (or substrate).

We also document a phonon Knudsen effect that causes a 
significant suppression of the in-plane thermal conductivity for 
nanoscale layer thicknesses at room temperatures and even 
micronscale thicknesses at liquid-nitrogen temperatures, 
$\theta\approx \hbox{77 K}$.  At the same time we show that 
the traditional Fourier analysis breaks down in the description 
of such small structures and stress that shifting toward a 
phonon based description of thermal dissipation will become 
increasingly important with a continued downscaling.

\vspace{\parskip}\textbf{2. Suppression of thermal transport across 
SiC/metal interfaces}

Figure 2 illustrates the atomistic simple-cubic (SC)
model adapted to estimate the finite-temperature thermal transport 
across the SiC/metal (or SiC/Si-substrate) interface. 
Both the phonon frequency $\omega$ and the in-plane 
momentum $q_{\|}$ (parallel to the interface) is 
conserved in this description of the phonon transmission.
While approximative, our model description retains the 
important transport consequences of phonon 
total-internal reflection~\cite{suplat,percond}.

This SC atomistic model has previously been used to calculate the 
phonon transmission and thermal transport in Si/Ge superlattices
and heterostructures~\cite{suplat,percond}.  Here we extend the SC
model to describe the binary nature of the SiC material by choosing 
the SC volume $a_{\rm sc}^{3}$ to contain exactly one physical 
atom, but having an average atomic mass $M_{\rm SiC} = (M_{\rm Si} 
+ M_{\rm C})/2$.  The value of $a_{\rm sc}=2.18$ {\AA} corresponds 
to the physical (FCC) lattice constant $a_0=4.36$ {\AA} of 
SiC~\cite{LB}.  Within each of the materials, these SC volumes 
are linked by force constants $F_{p,\,t;\,{\rm SiC}}$ 
with values which are fitted to the SiC sound 
velocities~\cite{LB} $c_{l,t;\,{\rm SiC}}=a_{\rm sc}
\sqrt{F_{p,t;\,{\rm SiC}}/M_{\rm SiC}}$, as expressed in 
the SC approximation~\cite{suplat}.  In this model we also 
obtain acoustic impedances~\cite{percond} $Z_{l,t;\,{\rm SiC}}=
\sqrt{F_{p,t;\,{\rm SiC}}M_{\rm SiC}}/a_{\rm sc}^2$.
We use a corresponding notation ($F_{p,t;\,{\rm me}}, 
M_{\rm me}, \ldots$) to describe the SC phonon dynamics 
in the neighboring media `me' (metal, Si, or general 
substrate). We assume for simplicity an identical lattice 
constant $a_{\rm sc}$ for both the SiC and the adjacent 
material, and we approximate the elastic coupling across the 
interface by $K_{p,t}\equiv
\sqrt{F_{p,t;\,{\rm SiC}}F_{p,t;\,{\rm me}}}$.

Our calculation of the phonon conductance $\sigma_{\rm K}$
across the SiC/metal or SiC/substrate interface follows the
approach and analysis of Ref.~\cite{percond} to which we 
refer for a more detailed discussion.  The modes in our SC 
model can be polarized either parallel (`p') or at right 
angles (`t') to the interface normal. In the following we 
focus on the description of the phonon dynamics and 
transmission of `p'-polarized modes.  The dynamics of a 
phonon moving in the SiC or neighboring media `me' 
is then unique characterized by the conserved set $(\omega,q_{\|})$.
We introduce frequencies $\Omega_{p,\,t;\,{\rm SiC\,(me)}} = 
2\sqrt{F_{p,\,t;\,{\rm SiC\,(me)}}/M_{\rm SiC\,(me)}}$ and the 
(conserved) dimensionless measure $\alpha_{q_{\|}}= 2-
\cos(q_x a_{\rm sc}) -\cos(q_y a_{\rm sc})$ of the in-plane 
dynamics. The SC-phonon dispersion relation can be expressed
\begin{equation}
\Omega_{p;\,{\rm SiC(me)}}^2
\left[1-\cos(k_{\rm SiC(me)} a_{\rm sc})\right]
=2\omega^2 -\Omega_{t;\,{\rm SiC(me)}}^2
\alpha_{q_{\|}}
\label{eq:atten}
\end{equation}
and serve to establish the perpendicular momentum component 
$k_{\rm SiC(me)}$ of any phonon mode $(\omega,q_{\|})$ 
when moving in the SiC (and/or in the metal or substrate `me').

We emphasize that this SC model description includes the affects 
of attenuation and of phonon total-internal reflection~\cite{suplat}. 
The attenuation (with associated thermal-transport suppression) 
arises because there is either insufficient or excess energy 
available for the perpendicular dynamics.  A phonon with 
in-plane momentum $q_{\|}>0$ that approaches from the typically
softer `me' side can be blocked from entering the hard SiC 
because the mode have too little energy to sustain the 
in-plane dynamics.  Conversely, a mode $(\omega,q_{\|})$ which 
propagates in the SiC (having a real value of $k_{\rm SiC}$) can
have too much energy to enter the (softer) metal or substrate. 
In our SC model, both of these situations causes the dispersion 
relation~(\ref{eq:atten}) to yield an imaginary value of 
$k_{\rm SiC(me)}$, {\it i.e.}, the correct attenuated 
behavior on the SiC (me) side of the interface.  At 
elevated temperatures and with a finite separation 
in the sound velocities (material hardness) such total-internal 
reflection suppresses the thermal transport contribution from 
most of the available phase space~\cite{suplat,percond}. 

To quantify the resulting suppression of the conductance
we calculate the SC estimates for the  probability 
$T_K(\omega,q_{\|})$ of phonon tunneling at general
$\omega$ and $q_{\|}$ within our SC model. 
We extract $T_K(\omega,q_{\|})$ by solving the Newton equation of 
motion for the phonon displacement $\eta_{q_{\|};\, {\rm SiC}}(l\leq 0)$ 
and $\eta_{q_{\|};\,{\rm me}}(s\geq 1)$ of the (atom-scale) SC 
elements located at the interface (here located between $l=0$ and $s=1$).
Specifically, we consider then the reflection and transmission 
of an incoming SiC mode
\begin{eqnarray}
\eta_{q_{\|};\,{\rm SiC}}(l\leq 0) & = &
e^{ik_{\rm SiC}a_{\rm sc} l} + A\, e^{-ik_{\rm SiC}
a_{\rm sc}l} 
\label{eq:dislp1}
\\
\eta_{q_{\|};\,{\rm me}}(s\geq 1) & = & B\,e^{ik_{\rm me}
a_{\rm sc}s},
\label{eq:dislp2}
\end{eqnarray}
and solve the harmonic equation
\begin{eqnarray}
\eta_{q_{\|};\,{\rm me}}(s=1) \,\,
\left(2\omega^2-\Omega_{t,\,{\rm me}}^2\alpha_{q_{\|}}\right)/2
& = &
\frac{F_{p;\,{\rm me}}}{M_{\rm me}}
\left[
\eta_{q_{\|};\,{\rm me}}(s=2)-\eta_{q_{\|};\,{\rm me}}(s=1)
\right]\nonumber \\
& + & \frac{K_{p}}{M_{\rm me}}
\left[
\eta_{q_{\|};\,{\rm me}}(s=1)
-
\eta_{q_{\|};\,{\rm SiC}}(l=0)
\right].
\label{eq:eqm}
\end{eqnarray}
The tunneling vanishes, of course, exactly when there is attenuation 
on either sides of the interface; in the case when both $k_{\rm SiC}$ 
and $k_{\rm me}$ are real, we determine the tunneling probability
\begin{equation}
T_K(\omega;\,\alpha_{q_{\|}}) = 1-{\left|A\right|}^2
\label{eq:transmission}
\end{equation}
and hence obtain a description of the transport contribution
at general $(\omega,q_{\|})$.

We calculate the finite-temperature thermal conductance
across the SiC/Metal interface by use of the formal 
result~\cite{Kapitza,percond}:
\begin{equation}
\sigma_K = \sum_{\rm modes} \int \frac{d^2q_{\|}}{{(2\pi)}^2}\,\,
{\left[\int 
\frac{d\omega}{2\pi}\; \hbar \omega \,T_K(\omega;\,q_{\|})
\left(\frac{\partial N_0}{\partial \theta}\right)
\right]}
\label{eq:cond}
\end{equation}
The `free' interface thermal conductance $\sigma_0$ that describe the
phonon transmission across a SiC/SiC interfaces (for example, 
between the doped SiC substrate and the undoped SiC transport 
channel) serves as a natural reference. We typically have the 
situation $\sigma_K(\theta) \ll \sigma_0(\theta)$ at finite
temperatures~\cite{percond} (for example, above the 
liquid-nitrogen temperature).

Figure 3 shows the results for the thermal conductance, $\sigma_K$, 
calculated at $\theta=\hbox{300 K}$ for a general SiC/metal or 
SiC/substrate interface within the SC model description. The
plot reports a dramatic variation of $\sigma_K$  as a function 
of the ratio of sound velocities $c_{\rm me}/c_{\rm SiC}$ and 
a much weaker dependence on the ratio of acoustic impedances 
$Z_{\rm me}/Z_{\rm SiC}$. Inserting the value for these ratios 
from Table I, the figure also documents that the typical 
SiC/metal and the SiC/Si interface at $\theta=\hbox{300 K}$ 
causes a dramatic suppression of the thermal conductance 
(when compared to the free conductance $\sigma_0$ value given 
by $c_{\rm me}/c_{\rm SiC}=Z_{\rm me}/Z_{\rm SiC}\equiv 1$).
The study identify the need for further improving our
understanding of the thermal coupling at such interfaces.

We also stress that this suppression of $\sigma_{\rm K}$ is 
qualitatively and quantitatively very different from the 
traditional conductance variation in low-temperature Kapitza 
effect. In the low-temperature regime the transmission is given by 
classical acoustics and hence exclusively defined by the ratio
$Z_{\rm me}/Z_{\rm SiC}$. In the present finite-temperature regime 
it is instead the total-internal phonon reflection and 
hence $c_{\rm me}/c_{\rm SiC}$  that specifies the dominant 
thermal-conductance variation.  In general the contact region 
between SiC and metals possesses a complex structure.  Our 
calculation of the $\sigma_K$ variation assumes a perfect 
junction between the SiC and the metal or Si substrate, rests 
on a simplified description of the characteristic phonon modes, 
and can thus only provide an approximative description. 
Nevertheless, our results strongly indicates that it is the
difference in sound velocities rather than (as often assumed)
the ratio of acoustic impedances which should be matched to maximize 
the high-temperature thermal transport out from a SiC power devices
at contacts, gates, and cooling fringes.

\vspace{\parskip}\textbf{3. Suppression of the thermal transport in SiC layers}

We calculate the phonon Knudsen effect on the in-plane
thermal transport by solving the linearized Boltzmann transport 
equation (l-BTE) in the relaxation-time approximation~\cite{boltzmann}.
We consider a layer of thickness $d_{\rm layer}$ in which a small 
thermal gradient, $\nabla \theta\|x$, induces a change 
$\delta N$ from the equilibrium phonon distribution $N_0= 1/( 
e^{\hbar \omega/k_B\theta}-1)$. The upper, lower materials boundaries 
are placed at $z_{>,<}=\pm d_{\rm layer}/2$ and we here use 
$q_{z}$/$q_{\|}$ to denote the component of the total phonon 
momentum, $q$, that is perpendicular to the layers/parallel to 
the in-plane temperature gradient $\nabla \theta$, 
respectively\footnote{The other in-plane momentum component 
is denoted $q_y$; we seek to keep a consistent notation between 
the two calculations although the direction of the temperature 
gradient differs.}. The bulk 
phonon mean free path, $l_{\rm mfp}={\bf v}_q \tau_q$, is given 
by a product of the phonon group velocity and the (bulk) relaxation 
time. We assume, for simplicity, that $l_{\rm mfp}$ is independent 
of the mode and momentum and we fit $l_{\rm mfp}$ against 
measurements of the bulk-SiC thermal conductivity~\cite{per1}. 
Solving then the l-BTE
\begin{eqnarray}
\left\{
\frac{q_{z}}{q}\, \frac{\partial}{\partial z} 
+\frac{1}{l_{\rm mfp}}\right\} \delta N
=-
\frac{1}{q}
\left(\frac{\partial N_0}
{\partial \theta}\right)
\left(q_\| \, |\nabla \theta| \right),
\label{eq:BTEgeneral}
\end{eqnarray}
in the absence and presence of the first term, we obtain
and compare the distribution changes, $\delta N_{\rm bulk}$ and
\begin{equation} 
\delta N_{\rm layer} = 
\delta N_{\rm Bulk}\left(1-h_{q_z}(z,\,p_+,\,p_-)\right).
\label{eq:gFunc}
\end{equation} 
These distribution changes, in turn, determine the phonon 
thermal conductivity
\begin{equation}
\kappa_{\rm bulk,layer} = -
{\left(\nabla \theta\right)}^{-1}
\sum_{\rm modes} 
\int \frac{d q_{\|}}{(2\pi)} 
\int \frac{d q_{z}}{(2\pi)}
\int \frac{d q_y}{(2\pi)}
\hbar\omega_q{\bf v}_q\,\,\delta N_{\rm bulk,layer}
\label{eq:heatcurrent}
\end{equation}
in the bulk and in the layered structure, respectively.

The interface scattering causes a suppression
\begin{equation} 
h_{q_z}(z,\,p_+,\,p_-) 
=\sum_{\xi=>,<}\phi_{q_z}^{\xi}(p_+,\,p_-)
e^{-|z-z_{\xi}|/\{l_{\rm mfp}(|q_{\bot}|/q)\}} 
\label{eq:hhFunc}
\end{equation} 
which depends on the fraction of specular $p_{+(-)}$ and diffusive 
$1-p_{+(-)}$ scattering at the upper (lower) material boundary. We 
determine the amplitudes $\phi_{q_z}^{>,<}$ of this variation 
from an analysis of the phonon scattering at the two interfaces, 
$z_{>,<}=\pm d_{\rm layer}/2$, following 
Refs.~\cite{per1,fuchs,broido}.  Specifically, for phonons 
that travel toward $z_>=d/2$ (with perpendicular momentum 
$q_z >0$), the appropriate boundary 
condition for specular scattering is $\delta N(\Omega,\,q_z) 
\equiv \delta N(\Omega,\,-q_z)$, while the corresponding 
boundary condition for diffusive scattering is 
$\delta N(\Omega,\,q_z<0) \equiv 0$.  A similar treatment 
yields the boundary conditions at the lower interface, $z_<$,
and an approach to determine the full variation of the 
suppression~(\ref{eq:hhFunc}) by interface scattering.

We introduce $\overline{p} = (p_++p_-)/2$ and $\hat{p} = 
\sqrt{p_+p_-}$ and express the resulting phonon Knudsen effect
\begin{equation} 
\frac{\kappa_{\rm layer}}{\kappa_{\rm Bulk}}(\eta,\,p_+,\,p_-) =
1 - \frac{3}{8\eta}
\left\{(1-\bar{p})-4\sum_{j=1}^\infty Q(j,\,\eta,\,\bar{p},\,\hat{p})
\right\}
\label{eq:Flayer}
\end{equation}  
using special functions
\begin{eqnarray} 
Q(j,\,\eta,\,\bar{p},\,\hat{p}) = 
(\hat{p}^2-2\bar{p}+1)\hat{p}^{j-1}D(j\eta)+(\bar{p}-\hat{p})(\hat{p}+1)^2
\hat{p}^{2(j-1)}D(2j\eta)
\label{eq:Qfunc}
\end{eqnarray}
and
\begin{eqnarray} 
D(a\eta) \equiv
\frac{1}{4}
\left(1-5\frac{(a\eta)}{3}-\frac{(a\eta)^2}{6}
+\frac{(a\eta)^3}{6}\right)e^{-a\eta}
+\frac{1}{2}
\left((a\eta)^2-\frac{(a\eta)^4}{12}\right)
\int_{\eta}^{\infty}dk\;\frac{e^{-ak}}{k}.
\label{eq:layer}
\end{eqnarray}
The analytical result extends earlier results~\cite{per1,broido}
and permits an efficient evaluation in the case where the 
degree of specular and diffusive scattering differs at the 
upper and lower boundary.  The result~(\ref{eq:layer}) describes,
in particular, the case when a thin (undoped) SiC transport 
channel is located below an oxide ($p_{+}=0$) and on top of 
a doped SiC substrate ($p_{-}\to 1$) or on a Si substrate 
($0< p_{-}< 1$, depending on the quality of 
the interface).

Figure 4 documents the significant phonon Knudsen effect caused
by the interface scattering in two typical structures (inserts)
at two different temperatures. The figure shows the variation in the
effective thermal conductivity $\kappa_{\rm eff}$ evaluated
as the weighted average of the (interface-limited) thermal
conductivities describing the (finite) oxide, (finite) SiC and 
(finite) Si-substrate layers. The figure documents that the 
Knudsen effects causes a very significant suppression of the 
thermal transport even up to hundreds of nanometers 
(micrometers) in the room (liquid-nitrogen) temperature cases.

Figure 4 further documents that calculations based on the
traditional approach using Fourier law of heat conductions
is inadequate due to the micron-scale phonon
mean free path. The panels compares calculations with such
a finite-element type description (dotted curves) against
the present phonon-based transport calculations 
assuming both diffusive (solid curves) and specular
(dashed curves) scattering at the SiC/Si-substrate
interface.  We find that the traditional approach 
completely breaks down at lower temperatures, and that
moving to phonon-based calculations of thermal transport
is essential for quantitative discussions of heat 
dissipation at room temperatures.

\vspace{\parskip}\textbf{4. Conclusions}

In this paper we have documented significant suppressions
of the thermal transport both across a SiC/metal or SiC/substrate 
interfaces and within the SiC layer between a top oxide and
the substrate.  For the room-temperature thermal conductance
we find that it is the ratio of material sound velocities,
rather than of the acoustic impedances, which should be
adjusted to maximize the thermal transport.  For the 
thermal transport in a layered structure we further 
documented that interface scattering causes a significant 
suppression for nanoscale (micronscale) layer thicknesses at 
room (liquid-nitrogen) temperatures.  Neither of these
heat-transport effects can be addressed within the traditional 
description of heat transport based on the Fourier law.  We conclude 
that the introduction of multiple interfaces and/or continued 
reduction of the thickness of the substrate or the conduction 
channels has adverse effects on the overall device heat dissipation.
The ongoing quest for miniaturization motivates a shift toward 
phonon-based calculations of the thermal transport in 
nanostructure devices.

\vspace{\parskip}\textbf{5. Acknowledgment}

This work was supported by the Swedish National Graduate School in 
Material Science (NFSM) and by the Swedish Foundation
for Strategic Research (SSF) through the consortium ATOMICS.


\pagebreak


\vspace{0.4\parskip}
\begin{center}
TABLES AND TABLE CAPTIONS
\end{center}

\begin{table*}[bc]
\begin{center}
\begin{tabular}{llcccccccccccc}
\hline
 &\multicolumn{10}{c}{}\\ 
& &  $a_0$ & M &  $a_{\rm sc}$ &  $c_l$ &$c_t$ & ${\left(\frac{Z_{\rm me}}{ Z_{\rm SiC}}\right)}_l$ 
& ${\left(\frac{Z_{\rm me}}{ Z_{\rm SiC}}\right)}_t$ &${\left(\frac{c_{\rm me}}{c_{\rm SiC}}\right)}_l$ &${\left(\frac{ c_{\rm me}}{ c_{\rm SiC}}\right)}_t$\\
&& \AA  & a.m.u & \AA &(m/s)&(m/s) && & & \\
\hline 
\\
& SiC(FCC) & 4.36 & 20.05 & 2.188 & 13035 & 6257 & 1 & 1 & 1 & 1 \\
& Si(FCC) & 5.43 & 28.09 & 2.715 & 8470 & 5340 & 0.65 & 0.85 & 0.48 & 0.63 \\
& Al(FCC) & 4.05 & 26.99 & 2.551  & 6360 & 3130 & 0.41 & 0.42 & 0.49 & 0.50\\
& Ni(FCC) & 3.52 & 58.71 & 2.218 & 5810 & 3080 & 1.25 & 1.38& 0.45& 0.49\\
& Cu(FCC) & 3.61 & 63.55 & 2.274  & 4760 & 2320 & 1.02 & 1.04 & 0.36 & 0.37\\
& Ag(FCC) & 4.09 & 107.87 & 2.576 & 3640 & 1690 & 0.92 & 0.89 & 0.28 & 0.27\\
\\
\hline
\end{tabular}
\end{center}
\end{table*}

\begin{center}
Table 1: Materials properties and model parameters for the calculation
of the room-temperature thermal transport across SiC/metal and/or 
SiC/Si interfaces.
\end{center}

\pagebreak

\begin{center}
FIGURES AND FIGURE CAPTIONS
\end{center}
\begin{figure}[h]
\begin{center}
\scalebox{0.58}{\includegraphics{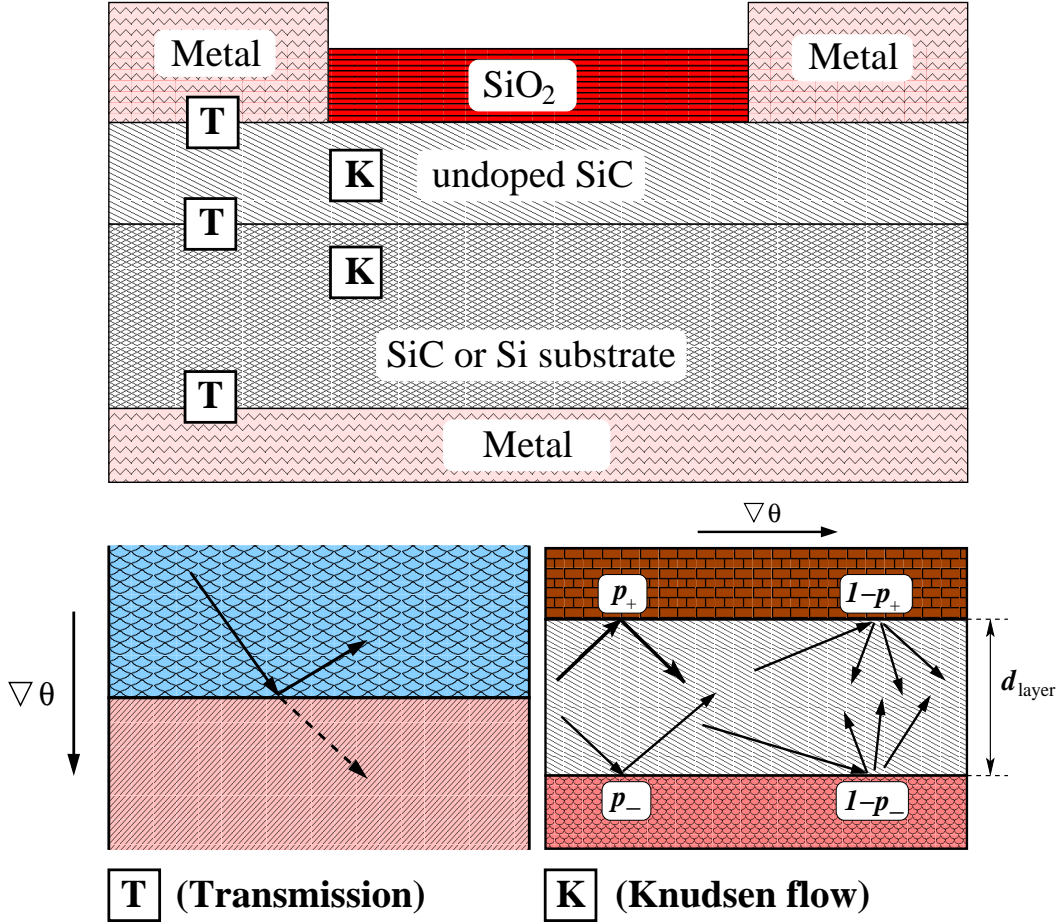}}
\caption{\label{MOS}
Structural schematics (top panel) and summary 
of key phonon transport effects (bottom panels) in a typical 
nanoscale SiC device.  Oxide and substrate layers together with 
SiC/metal interfaces surrounds the thin (undoped) SiC transport 
channel.  We find that heat dissipation is suppressed at room 
temperatures by the ineffective phonon transmission `T' (lower 
left panel) across the SiC/metal or SiC/Si interfaces {\it 
and\/} by the phonon Knudsen effect `K' (lower 
right panel), that is, the suppression of the thermal
conductivity within the SiC by interface scattering at 
layer thicknesses $d_{\rm layer} < \hbox{0.5 $\mu$m}$.  The 
phonon transport is calculated analytically at general 
probabilities, $p_{+(-)}$ and $1-p_{+(-)}$, for specular 
and diffusive scattering at the upper (lower) interface.}
\end{center}
\end{figure}

\pagebreak
                                                                                
\vspace{1.4\parskip}

\begin{figure}[c]
\begin{center}
\scalebox{0.62}{\includegraphics*{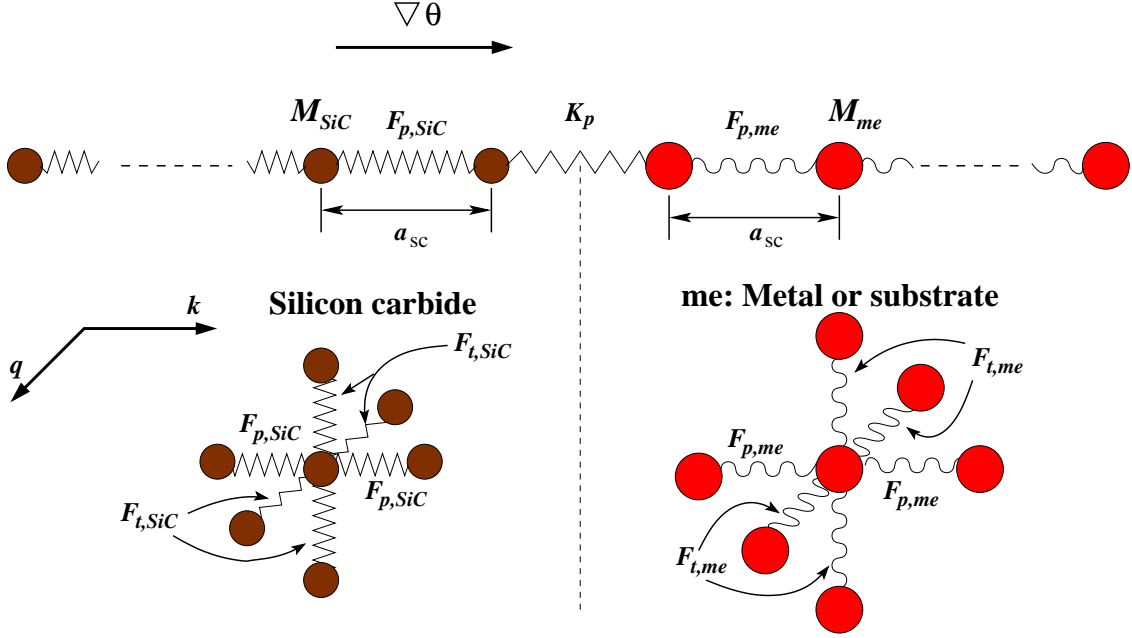}}
\caption{\label{spring}
Schematics of the effective, atomistic 
representation in a simple-cubic (SC) model~\protect\cite{suplat} 
adapted to estimate the finite-temperature phonon transport 
across SiC/metal (and SiC/Si) interface (vertical dashed line).  
The SC volume $a_{\rm sc}^{3}$ is chosen to contain exactly one 
physical atom but having the average atomic mass $M_{\rm SiC} 
= (M_{\rm Si} + M_{\rm C})/2$. A similar approach describes
the neighboring media `me' (metal or Si substrate).  Force 
constants, $F_{p,t;\,{\rm SiC\,(me)}}$ fitted to the longitudinal 
and transverse sound velocities~\protect\cite{suplat,percond}, 
link the SC volumes within the SiC (metal) while elastic 
coupling constants $K_{p,t} \equiv \protect\sqrt{F_{p,t;\,
{\rm SiC}}F_{p,t;\,{\rm me}}}$ connect across the 
interface~\protect\cite{percond}.}                                                                                
\end{center}
\end{figure}

\pagebreak
                                                                                
\vspace{1.4\parskip}
\begin{figure}[c]
\begin{center}
\scalebox{0.8}{\includegraphics{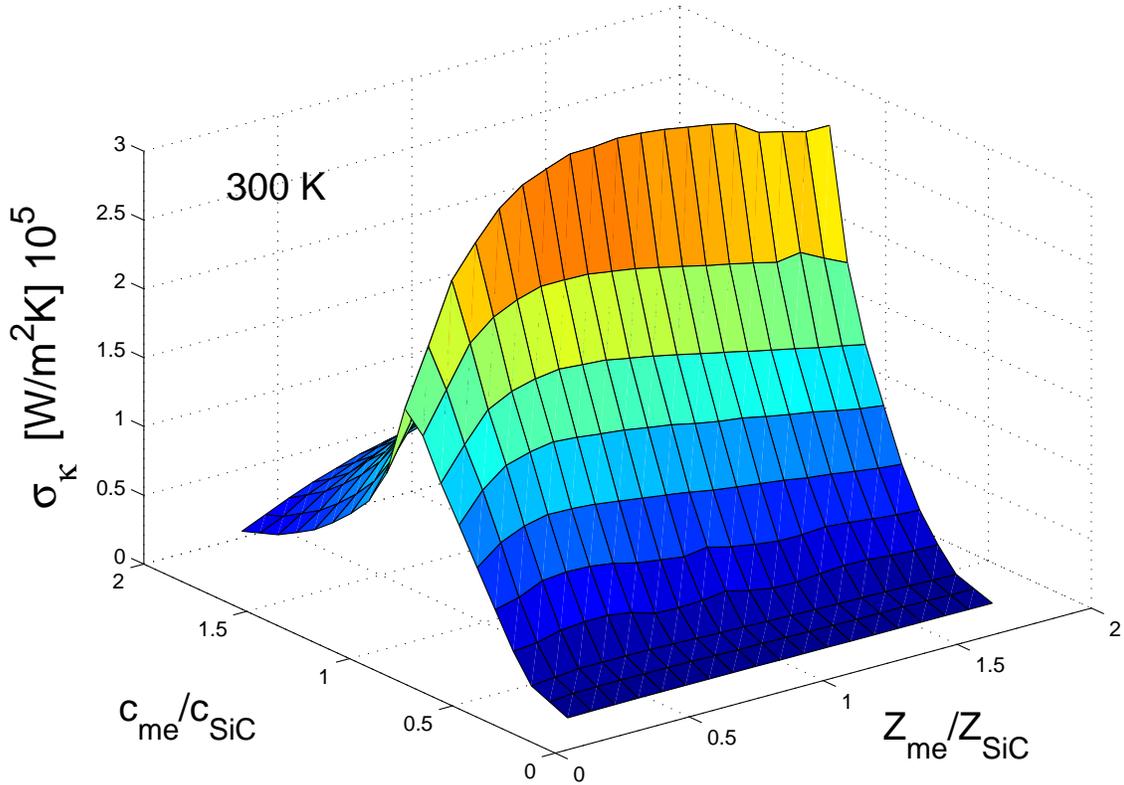}}
\caption{\label{Kapitza}
Dramatic variation of the room-temperature
thermal conductance $\sigma_K$ across a SiC/general-media 
interface specified by key differences in materials properties.
A strong (weaker) dependence arises with the variation
in the ratio of material sound velocities $c$ (acoustic 
impedance $Z$) which specifies the amount of total internal
phonon reflection (long-wavelength tunneling probability).}
\end{center}
\end{figure}

\pagebreak
                                                                                
\vspace{1.4\parskip}

\begin{figure}[h]
\begin{center}
\scalebox{0.8}{\includegraphics{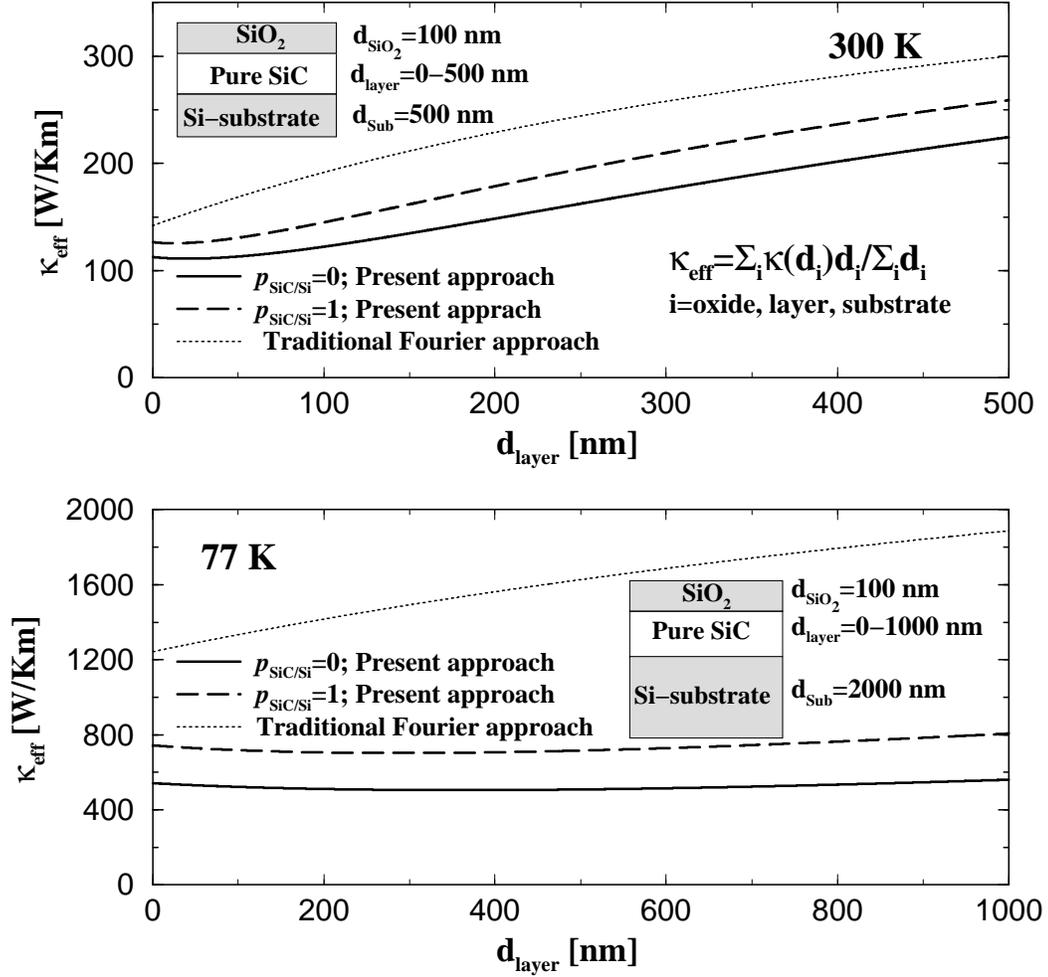}}
\caption{\label{Knudsen_Suppression}
Knudsen suppression of the effective in-plane SiC 
transport $\kappa_{\rm eff}$ by interface scattering in 
the typical layered structures (inserts). The upper (lower)
panel reports the variation with the thickness $d_{\rm layer}$
of SiC conductance channel at room (liquid-nitrogen) 
temperatures assuming either diffusive (solid curves) or specular
(dashed curves) scattering at the SiC/Si-substrate interface. The 
traditional approach based on the Fourier law of heat flow (dotted 
curves) is documented to be inadequate due to the 
micron-scale phonon mean-free path.}
\end{center}
\end{figure}

\end{document}